\documentclass[12pt,aps,tightenlines,notitlepage,superscriptaddress]{revtex4-1}
\usepackage{amsmath}
\usepackage{amssymb}
\usepackage{amsfonts}
\usepackage{bm}
\usepackage[mathscr]{eucal}
\usepackage{theorem}
\usepackage{graphicx}
\usepackage{psfrag}
\usepackage{subfigure}
\usepackage{color}
\usepackage{wasysym}
\include{graphix}
\usepackage{framed}
\usepackage{enumitem}%
\usepackage{lipsum}
\usepackage{float}
\usepackage{MnSymbol}
\usepackage{url}
\usepackage{tikz}
\usepackage[color=yellow]{todonotes}
\usetikzlibrary{calc}
\usepackage[colorlinks]{hyperref}

\graphicspath{{figs/}}

\usepackage{geometry}
\newcommand{\myqed}{\nobreak \ifvmode \relax \else
      \ifdim\lastskip<1.5em \hskip-\lastskip
      \hskip1.5em plus0em minus0.5em \fi \nobreak
      \vrule height0.75em width0.5em depth0.25em\fi}

\newcommand{\rem}[1]{}
\newcommand{\xcl}{x_0}
\newcommand{\xmac}{x_{\mathrm{m}}}

\newcommand{\mathd}{\mathrm{d}}
\newcommand{\mathe}{\mathrm{e}}
\newcommand{\imag}{\mathrm{i}}
\newcommand{\myhbar}{\overline{h}}
\newcommand{\Diff}{\mathcal{D}}
\newcommand{\vech}{\bm{\overline{h}}}

\begin{document}
\title{A Geometric Diffuse-Interface Method for Droplet Spreading}

\author{Darryl D. Holm}
\email{d.holm@ic.ac.uk}
\affiliation{Department of Mathematics, Imperial College London, London SW7 2AZ, United Kingdom}

\author{Lennon \'O N\'araigh}
\email{onaraigh@maths.ucd.ie}
\affiliation{School of Mathematics and Statistics, University College Dublin, Belfield, Dublin 4, Ireland}

\author{Cesare Tronci}
\email{c.tronci@surrey.ac.uk}
\affiliation{Department of Mathematics, University of Surrey, Guildford GU2 7XH, United Kingdom}
\affiliation{Numerical Methods Division, Max Planck Institute for Plasma Physics, Garching 85748, Germany}

\date{\today}

\begin{abstract}
This paper exploits the theory of geometric gradient flows to introduce an alternative regularization of the thin-film equation. The solution properties of this regularization are investigated via a sequence of numerical simulations whose results lead to new perspectives on thin-film behavior. The new perspectives in large-scale droplet-spreading dynamics are elucidated by comparing numerical-simulation results for the solution properties of the current model with corresponding known properties of three different alternative models. The three specific comparisons in solution behavior are made with the slip model, the precursor-film method and the diffuse-interface model. 
\end{abstract}

\maketitle

\section{Introduction}

\begin{figure}
\centering
\includegraphics[width=0.8\textwidth]{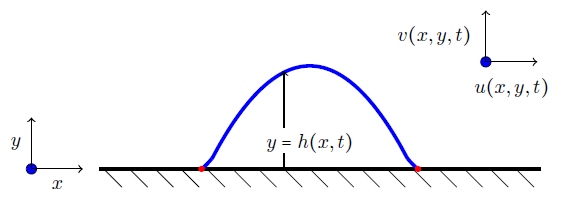}
\caption{Schematic description of the fluid mechanical problem of droplet spreading, as derived from the Navier--Stokes equations in the lubrication limit}
\label{fig:sketch1}
\end{figure}

This work is concerned with the thin-film equation
\begin{subequations}
\begin{equation}
\frac{\partial h}{\partial t}=-\frac{\partial}{\partial x}\left(h^n\frac{\partial^3 h}{\partial x^3}\right),\qquad t>0,\qquad x\in (-\infty,\infty),
\end{equation}
\begin{equation}
h(x,t=0)=h_0(x),\qquad h_0(x)\geq 0,\qquad x\in(-\infty,\infty),
\end{equation}%
\label{eq:thif_basic}%
\end{subequations}%
and its modifications.  Boundary conditions are chosen as $|x|\rightarrow \infty$ so that 
Equation~\eqref{eq:thif_basic} conserves mass:
\begin{equation}
\frac{\mathd M}{\mathd t}=0,\qquad M=\int_{-\infty}^\infty h(x,t)\mathd x.
\label{eq:mass}
\end{equation}
 The particular value $n=3$ is physically relevant, as then Equation~\eqref{eq:thif_basic} is a model for the free surface of a viscous thin-film flow.  Indeed,  equation~\eqref{eq:thif_basic} with $n=3$ amounts to the Navier--Stokes equations for a thin-film flow, in the limit of lubrication theory.  This derivation is developed in Reference~\cite{oron1997long}. A sketch of the physical scenario is given in Figure~\ref{fig:sketch1}.
The case $n=3$ is the subject of the present article. In particular, we revisit the problem of \textit{droplet spreading}, that is, we seek to model the time evolution of an initial profile $h_0(x)$ with compact support. 

For a general value of $n$, the spreading of droplets governed by Equation~\eqref{eq:thif_basic} admits the similarity solution~\cite{bertozzi1996lubrication}
\begin{equation}
h(x,t)=t^{a}f(x/t^a),\qquad a=\frac{1}{n+4}.
\label{eq:similarity}
\end{equation}
Substituting this trial solution into Equation~\eqref{eq:thif_basic} yields the ordinary differential equation
\begin{equation}
f^n f'''=\frac{nf}{n+4}.
\label{eq:f_eqn}
\end{equation} 
For $n<3$ Equation~\eqref{eq:f_eqn} possesses smooth solutions with compact support.  In this case, equation~\eqref{eq:f_eqn} with initial conditions $f(0)=1$, $f'(0)=0$, and $f''(0)=-\mu<0$,  can be solved by adjusting $\mu$ until $f=f'=0$ at some $\eta=\eta_0>0$, corresponding to the outermost extent of the droplet.  Thus, the position $\xcl$ of the microscopic contact line where the free surface $h(x,t)$ touches down to zero is described by $\xcl=\eta_0 t^{1/(n+4)}$.  Unfortunately, this description breaks down precisely for the physically relevant value of $n=3$, at which $f(\eta)$ degenerates into a Dirac delta function centred at $\eta=0$, and the droplet does not spread.

Physically, the breakdown of Equation~\eqref{eq:thif_basic} as a model of droplet spreading for $n=3$ is due to a small but not ignorable effect which occurs in the vicinity of the microscopic contact line.  Namely, the modelling assumptions which enable the passage from the Navier--Stokes equations for a thin film to the simplified free-surface evolution equation~\eqref{eq:thif_basic} assume there is no relative motion between the film and the underlying substrate. (This is the no-slip condition.)  However, the no-slip condition is not consistent with a moving contact line.  Many different  approaches have been proposed in the literature to restore the missing physics. Three of these approaches will be summarised below.  These approaches all exhibit the same qualitative behaviour. However, they each have certain drawbacks. The short summaries of the three main approaches given below will provide the context in which our own proposal for healing the contact-line singularity will be introduced. 

{\textbf{Slip-length modelling:}} A common approach in the modelling literature to resolving the paradox of the moving contact-line is  to modify Equation~\eqref{eq:thif_basic} (with $n=3$) as follows,
\begin{equation}
\frac{\partial h}{\partial t}=-\frac{\partial}{\partial x}\left[\left(h^3+\lambda h^2\right)\frac{\partial^3 h}{\partial x^3}\right],\qquad t>0,\qquad x\in (-\infty,\infty),
\label{eq:slip}
\end{equation}
where $\lambda$ is a positive (dimensionless) constant related to the slip length.    Equation~\eqref{eq:slip} can be derived using lubrication theory, starting from the Navier--Stokes equations.  Instead of imposing the no-slip condition on the velocity component $u(x,y=0,t)$ tangent to the substrate, one instead imposes the condition 
\begin{equation}
u(x,y=0,t)=\lambda_* \left(\frac{\partial u}{\partial y}\right)_{y=0},
\label{eq:navier}
\end{equation}
where $\lambda_*$ is the dimensional slip length.  Working in the limit of lubrication theory, and using appropriate non-dimensionalization, one obtains Equation~\eqref{eq:slip} in this manner.

\begin{figure}
\centering
		\includegraphics[width=0.8\textwidth]{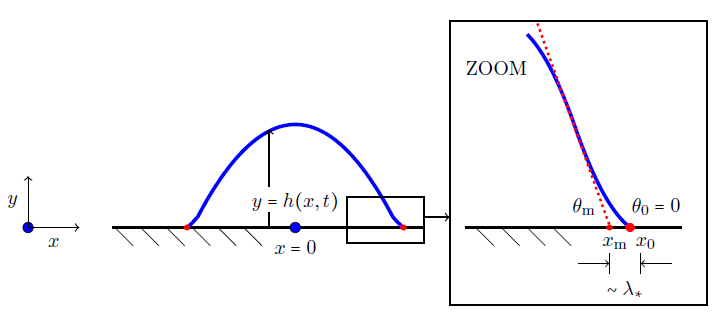}
\caption{Schematic description of the fluid mechanical problem of droplet spreading, showing the inner and outer regions of the problem.}
\label{fig:sketch2}
\end{figure}

Typically, one works with $\lambda \ll 1$, as the effect of slip is a small but not ignorable.  Then, Equation~\eqref{eq:slip} can be solved via the method of matched asymptotic expansions~\cite{eggers2004characteristic}.  In this approach, one distinguishes between inner and outer solutions, separated by a lengthscale $\xmac(t)$ which demarcates the regions of validity of the different solutions.  The geometry of this setup is shown in Figure~\ref{fig:sketch2}.  
In the outer region, with $|x|\ll \xmac(t)$, the effect of slip is ignorable, and one can work with $\lambda=0$ in Equation~\eqref{eq:slip}.  The free-surface profile in this outer region can therefore be well approximated by the similarity solution~\eqref{eq:f_eqn} with $n=3$.  By not continuing this solution past $|x|=\xmac$, the singularity that occurs in the similarity solution is avoided.  

In the inner region, we identify the microscopic contact line $x=\xcl(t)$ where the free-surface height touches down to zero in a smooth fashion, $h(\xcl,t)=0$ and $h_x(\xcl,t)=0$.  As such, the inner region corresponds to
 $|x-\xcl|\ll \lambda$.  In this region,  one  therefore  solves Equation~\eqref{eq:slip} without the $h^3$ term.  It is the disappearance of the $h^3$ term in this limit which enables the smooth touchdown of the solution at $x=\xcl(t)$.    Finally, the inner and outer solutions are matched at the scale $\xmac$.  As such, the position $\xmac$ is interpreted as the macroscopic contact line, and $\tan\theta_{\mathrm{m}}=h_x(\xmac,t)$ has the interpretation of the macroscopic contact angle.  Correspondingly, the position $\xcl$ has the interpretation of the microscopic contact line, meaning that the microscopic contact angle $\tan\theta_0=h_x(\xcl,t)$ is zero.  It can be noted that this description deals with perfect wetting, such that the droplet spreads indefinitely, and the macroscopic contact angle never reaches a constant value.

Up to prefactors, the matched-asymptotic expansion procedure produced above gives the following relation for the macroscopic contact angle:
\begin{equation}
\left(\frac{\partial h}{\partial x}\right)_{x=\xmac}^3 \sim \left(\frac{\mathd \xmac}{\mathd t}\right)\ln\left[\frac{\xmac}{\lambda b}\left(\frac{\mathd \xmac}{\mathd t}\right)^{1/3}\right],
\label{eq:voinov}
\end{equation}
where $b$ is a constant. (Equation~\eqref{eq:voinov} was also discovered via a different approach, in Voinov~\cite{Voinov1977}.)  Upon identifying the outer solution on the left-hand side of Equation~\eqref{eq:voinov} with the similarity solution~\eqref{eq:similarity} $h=t^{-a}h(x/t^a)$ with $a=1/7$, Equation~\eqref{eq:voinov} reduces to
\[
\left(\frac{\mathd \xmac}{\mathd t}\right) t^{6/7}=\text{Const.}+\text{Logarithmic corrections},
\]
hence, the leading-order behaviour of Equation~\eqref{eq:voinov} is given by $\xmac \sim t^{1/7}$, which is the experimentally validated Tanner's Law~\cite{oron1997long}, valid for a droplet spreading on a perfectly hydrophobic substrate (`complete wetting').  
As such, the leading-order behaviour of the Navier slip model is consistent with experimental findings. 

Although the Navier slip model~\eqref{eq:navier} alleviates the singularity in the free-surface height $h(x,t)$ at the microscopic contact line, the higher spatial derivatives of $h(x,t)$ remain singular there.  This means the capillary pressure $P=-h_{xx}$ is singular at the microscopic contact line.  Although the resulting singularity is rather mild, it does correspond to infinite pressure, which is undesirable in a physical model  (although the integral of the pressure, or the force, does remain finite).  Also, the prescription of the slip model~\eqref{eq:navier}, while convenient from the modelling point of view, does not have an \textit{a priori} theoretical basis, beyond the obvious connection to atomic-scale fluid-substrate interactions. These two drawbacks have motivated  the development of other models of droplet spreading.

{\textbf{Precursor-film modelling:}}    Here, a potential function is prescribed which governs the molecular interactions  between the fluid particles and the substrate.  
Typically, the chosen model potential function has the form of a Van der Waals potential with attractive and repulsive components~\cite{oron1997long}. This model is motivated by a solid understanding of the fluid-substrate interaction.  In the lubrication limit, the result of this modelling step is again a single equation for the droplet profile $h(x,t)$; instead of $h_t=-\partial_{x}(h^3 \partial_x P)$ with $P=-h_{xx}$ one has $P=-h_{xx}+\Phi(h)$, where $\Phi$ is fluid-substrate interaction potential~\cite{o2002theory}.  This description gives a solution for $h(x,t)$ which allows for droplet spreading.  Furthermore, the model allows for the possibility of equilibration of the spreading process whereby the film assumes a constant shape characterized by an equilibrium contact angle, thereby modelling partial wetting.  Crucially, the value of the equilibrium contact angle is expressed in terms of the coefficients in the Van der Waals potential; hence, the model is fully characterized by basic physical considerations.
A further advantage of the model is  the droplet-spreading solution maintains a finite stress at the contact line -- and yet the leading-order behaviour of the precursor-film and Navier-slip models agree, in a theory of matched asymptotic expansions~\cite{sui2014numerical}.  However, a drawback of the model is the requirement for the droplet solution to possess a precursor film -- an ultra-thin but non-vanishing film that stretches out indefinitely beyond the droplet core, whose thickness is set by a balance between the attractive and repulsive components in the potential $\Phi$.  It can be readily understood that this feature is undesirable in certain applications -- for instance, in drop deposition on a substrate, where the substrate should properly be assumed to be initially free from contamination by any fluid film (no matter how thin). 

\textbf{{Diffuse-interface modelling:}} We summarise the diffuse-interface model in the context of the full Navier--Stokes equations, from which the thin-film equation~\eqref{eq:thif_basic} emerges as a limit.  In the full Navier--Stokes equations, the motion of a droplet over a wall can be modelled using such a method.   In addition to the velocity and pressure fields, an auxiliary order-parameter variable $C$ (the `phase field') is introduced, which tracks the phases, such that $C=1$ inside the droplet, and $C=0$ outside, in the surrounding phase.
 There is a smooth transition between these two extreme values across a finite width -- hence, a diffuse interface.  The phase field evolves according to its own evolution equation, which is typically taken as the Cahn--Hilliard equation~\cite{ding2007diffuse}, thereby resulting in a mathematically consistent framework.
%
By introducing the diffuse interface in this manner (rather than having a sharp interface at the contact line), the model effectively produces slip, through the diffusive fluxes. Hence, the stress singularity at the moving contact line is removed even when a no-slip velocity boundary condition is imposed~\cite{ding2007wetting}.

In this work, we propose a regularisation of Equation~\eqref{eq:thif_basic} in the spirit of the diffuse-interface method.  As such, we propose a modified version of Equation~\eqref{eq:thif_basic} which depends not only on the free-surface height $h(x,t)$, but also on a diffuse free-surface height,
\begin{equation}
\myhbar(x,t)=\int_{-\infty}^\infty K(x-y;\alpha)h(y,t)\mathd y:=K*h,
\label{eq:filter}
\end{equation}
where $K(s;\alpha)$ is a smoothing kernel which smooths out small-scale features on a lengthscale $\alpha$ or less.  The regularisation is not ad-hoc, instead, it is introduced in the context of a rigorous gradient-energy theory in Section~\ref{sec:theory}.  The proposed regularisation method has some advantages over the other methods considered herein, in particular:
\begin{itemize}[noitemsep]
\item In contrast to the \textit{slip model}, the proposed regularisation produces a continuous pressure profile everywhere.  We demonstrate below that our new theoretical model amounts to imposing the usual fluid-mechanical interfacial conditions on the filtered free-surface height $\myhbar$, rather than on $h$.  In a context where the free surface $h$ comes into contact with a substrate which has some atomic level of roughness, this makes physical sense, and the use of $\myhbar$ reflects our uncertain knowledge of the precise location where the droplet, the surrounding medium, and the substrate all come into contact.  The parameter $\alpha$ can therefore be viewed as expressing this uncertainty.
\item In contrast to the \textit{precursor-film method}, we do not require a precursor film of small-but-finite thickness to extend to infinity.  Our model effectively has a precursor film whose thickness falls off to zero at large distances from the droplet core, the falloff scale is exactly $\alpha$.
\item Although motivated by the \textit{diffuse-interface concept}, our model does not dispense with the classical description of a sharp interface separating the fluid phases.  As such, the sharp interface is still contained in our model; although it is no longer a dynamic variable, and it is recovered from what is effectively a diffuse interface via a deconvolution operation (\textit{cf}. Equation~\eqref{eq:filter}).
\end{itemize}
By now one may have recognised that each of the above canonical methods involves a small length scale, as summarised in Table~\ref{tab:length scales}.  Therefore, it can be seen that each of the methods is concerned effectively with parametrisation missing the small-scale physics, to produce the same large-scale droplet-spreading dynamics in each case.    Our model also fits into this framework, as shown in Table~\ref{tab:length scales}.  We will further illustrate this and other features of our model by using numerical simulations in Section~\ref{sec:numerics} below. 
%
%
\begin{table}
\caption{Summary of the various small length scales used in the different regularisation methods}
	\centering
		\begin{tabular}{|c|c|}
		\hline
		Method & Lengthscale \\
		\hline
		Navier Slip Model & Slip length \\
		Attractive / Repulsive Potential & Precursor-film thickness \\
		Diffuse-Interface Method & Diffuse-interface thickness \\
		The present method & Length scale of smoothing kernel \\
		\hline
		\end{tabular}
\vspace*{-4pt}
\label{tab:length scales}
\end{table}

\section{Geometric Gradient-Flow structure}
\label{sec:theory}

The starting-point of the theoretical analysis is to notice that under the dynamics of Equation~\eqref{eq:thif_basic} the following free-energy functional decays with time:
\begin{equation}
E=\tfrac{1}{2}\int_{-\infty}^\infty |\partial_x h|^2\mathd x.
\label{eq:energy}
\end{equation}
Indeed, since $\delta E/\delta h=-\partial_{xx}h$, Equation~\eqref{eq:thif_basic} can be re-written as
\begin{equation}
\frac{\partial h}{\partial t}=\frac{\partial }{\partial x}\left[h \mu(h)\frac{\partial}{\partial x}\frac{\delta E}{\delta h}\right],\qquad \mu(h)=h^2.
\label{eq:energy1}
\end{equation}
By multiplying both sides of Equation~\eqref{eq:energy1} by $\delta E/\delta h$ and integrating from $x=-\infty$ to $x=\infty$, one obtains
\begin{equation}
\frac{\mathd E}{\mathd t}=-\int_{-\infty}^{\infty} h \mu(h)|\partial_{xx}h|^2\mathd x\leq 0.
\label{eq:energy3}
\end{equation}
The proposal for the regularized version of Equation~\eqref{eq:energy} is
\begin{equation}\label{eq:energyreg}
\bar{E}=\tfrac{1}{2}\int_{-\infty}^\infty |\partial_x\myhbar|^2\mathd x,
\end{equation}
where $\myhbar$ is the filtered free-surface height given by Equation~\eqref{eq:filter}.
Then, 
\[
\frac{\delta \bar{E}}{\delta h}=K*\frac{\delta {E}}{\delta \myhbar}=-K*\partial_{xx}\myhbar,
\]
so that equation \eqref{eq:energy1} becomes
\begin{equation}
\partial_t h=-\,\partial_x\left[h\overline{\mu}(h,\myhbar)\partial_x K*\partial_{xx}\myhbar\right],
\label{eq:h_reg1}
\end{equation}
where we have defined $\overline{\mu}(h,\myhbar)$ so that the mobility depends in general on both $h$ and $\myhbar$.

Equation \eqref{eq:energy1} belongs to a class of geometric gradient-flow equations which appeared in \cite{HP2,HP3,HPT1}. These gradient flows were first inspired by Darcy's Law for highly viscous flows, which establishes a proportionality relation between the velocity and the force experienced by the fluid. This construction generally applies to arbitrary tensor fields on the configuration manifold $M$ and it involves concepts of Geometric Mechanics, such as Lie derivatives and momentum maps. More specifically, if $V$ denotes the space of  tensor fields and $V^*$ its dual, one usually considers the duality pairing $\langle\cdot,\cdot\rangle:V^*\times V\to \mathbb{R}$ given by the standard $L^2-$pairing. This pairing can be used to define a momentum map $\diamond:V^*\times V\to \mathfrak{X}(M)^*$ whose target space $\mathfrak{X}(M)^*$ is identified with the space of one form-densities on $M$, that is $\mathfrak{X}(M)^*=\Lambda^1(M)\otimes\operatorname{Den}(M)$. In practice, upon defining by $\mathfrak{X}(M)$ the space of vector fields on $M$, this momentum map is defined as
\begin{equation}\label{momap}
\langle\,\zeta \diamond \nu\,,\, u  \,\rangle
:= \langle\,\zeta \,,\, -\pounds_u  \nu  \,\rangle\,,
\end{equation}
for any $u\in\mathfrak{X}(M)$ and any couple $(\nu,\zeta)\in V\times V^*$. Here, $\pounds_u$ denotes the Lie derivative with respect to $u$ and we use the $L^2-$pairing on both sides of the equality. Further, we assume that $M$ is a Riemmanniann manifold so that one can define the musical isomorphism  $(\,\cdot\,)^\flat: \mathfrak{X}\to \mathfrak{X}^*$ (flat) and its inverse $(\,\cdot\,)^\sharp: \mathfrak{X}^*\to \mathfrak{X}$ (sharp). In terms of these operations, a {\it geometric gradient-flow} on $V$ is given by an equation of motion of the type
\begin{equation}\label{GOP}
a_t = - \pounds_{\left( \mu(a)\diamond\frac{\delta E}{\delta a}\right)^\sharp\,} a
\,,
\end{equation}
where $\mu: V\to V^*$ is a {\it generalized mobility} and $E=E(a)$ is the energy functional, whose functional derivative is denoted by $\delta E/\delta a$. The geometric equation \eqref{GOP} has the following variational formulation \cite{HP3,HPT1}, which unfolds its gradient-flow nature: for an arbitrary $\zeta\in V^*$, one writes
\[
\left\langle\zeta,a_t\right\rangle=\left\langle\frac{\delta E}{\delta a},\delta a\right\rangle
\,,\qquad\text{ with } \qquad
\delta a = - \pounds_{\left( a\diamond\zeta\right)^\sharp\,} \mu(a)
\,.
\]
It is easy to see that Equation~\eqref{GOP} follows from the above by integrating by parts and using  the definition~\eqref{momap}.

Equation \eqref{eq:energy1} belongs to the class of geometric gradient-flow equations \eqref{GOP}. This may be shown as follows. Let $V=\operatorname{Den}(\Bbb{R})$, so that $\pounds_u a = \partial_x(u a)$ and $\zeta \diamond a=a\partial_x \zeta$. Then, the sharp operator becomes trivial and equation \eqref{GOP} reduces to  \eqref{eq:energy1} for $a=h{\rm d} x$. Notice that in the case of equation \eqref{eq:h_reg1}, we have extended the notion of {generalized mobility} such that $\mu: V\times V\to V^*$.
In this case,  equation \eqref{eq:h_reg1} is associated to the regularized version \eqref{eq:energyreg} of the energy functional \eqref{eq:energy}. Interestingly enough, the latter belongs to the  Burbea-Rao class \cite{BR1,BR2} of information norms on probability densities. More specifically, the energy functional \eqref{eq:energy} identifies the norm associated to a {\it $2^{\rm nd}-$order entropy metric} in the Burbea-Rao class. We shall leave this connection to information geometry as a direction for future studies.

An important property of geometric-gradient flows of the type \eqref{GOP} is that, when the generalized mobility and the functional derivative $\delta E/\delta a$ are sufficiently smooth, equation \eqref{GOP} admits singular solutions of the type \cite{HP3,HPT1}
\[
a(x,t)=\sum_{i=1}^N\alpha_i(t)\delta(x-q_i(t))
\,,\qquad\text{ with }\qquad
\dot{q}_i=\left(\mu(a)\diamond\frac{\delta E}{\delta a}\right)\bigg|_{x=q_i}
\,.
\]
The dynamics of the weights $\alpha_i(t)$ can be found on a case-by-case basis by direct substitution. In the case of equation \eqref{eq:h_reg1}, the existence of these solutions depends on the specific expression of $\bar\mu(h,\bar{h})$. If this is smooth enough after replacing the singular solution ansatz, then one easily verifies that the weights $\alpha_i$ are all constant and
\begin{equation}\label{singsol}
\dot{q}_i(t) = \Big[\bar\mu(h,\bar{h}) \partial_x\big(\bar{h}-\bar{\bar{h}}\big) \Big]_{x=q_i(t)} 
\,,
\end{equation}
where we have denoted $\bar{\bar{h}}=K*(K* h)$. We make three remarks about equation~\eqref{singsol}:
\begin{enumerate}[noitemsep]
\item $\bar\mu(h,\bar{h})$ is a function(al) of both ${h}(x,t)=\sum_{j} \alpha_j \delta(x-q_j(t))$ and $\bar{h}(x,t)=\sum_{k} \alpha_k K(x-q_k(t))$; 
\item $(\bar{h}-\bar{\bar{h}})$ is a function of $(x-q_j(t))$ and constant weights $\alpha_j$, summed over  indices $j$;  
\item After the functional dependence of the mobility $\bar\mu$ has been specified, the $x$-dependences of $(\bar{h}-\bar{\bar{h}})(x-q_j(t))$ and $\bar\mu(h,\bar{h})$ are both evaluated at $x=q_i(t)$ to produce a closed dynamical system for the positions $q_i(t)$ for each $i=1,2,\dots,N$.
\end{enumerate}
The singular solutions \eqref{singsol} exist, provided $\delta E/\delta h$ is a smooth functional derivative, which holds for our previous energy functional \eqref{eq:energyreg}. However, the singular solutions also require  a smooth generalized mobility. Indeed, the above notation $\bar\mu(h,\bar{h})$ is suggestive of an extra smoothing possibly occurring in the mobility function(al). For example, given a mobility function $\mu(h,\bar{h})$, a smooth mobility may be introduced by writing $\bar\mu=K*\mu(h,\bar{h})$. In certain cases, previous work has shown \cite{HP1,HP2,HP3} that the singular solutions of geometric gradient-flow equations emerge spontaneously from arbitrary smooth initial conditions and this behavior was exploited to model self-aggregation and alignment of particles with anisotropic interactions \cite{HP3,HPT2,HPT4,HONT4}. 

In the present work, we shall take an alternative route. Instead of studying singular solution dynamics, we exploit the  construction of geometric gradient-flows to introduce a new regularization of the thin-film equation. Then, we shall focus on studying the properties of this regularization without inserting any extra smoothing into the generalized mobility $\bar{\mu}(h,\bar{h})$. While this case precludes the existence of the singular solutions \eqref{singsol}, it still leads to new perspectives on thin-film behavior. Here, we choose the functional form
\begin{equation}
\overline{\mu}(h,\myhbar)=\tfrac{3}{2}h\myhbar-\tfrac{1}{2}h^2.
\label{eq:mob}
\end{equation}
Equation~\eqref{eq:h_reg1} with the mobility~\eqref{eq:mob} also has a physical basis: it can be obtained from the Navier--Stokes equations and lubrication theory by imposing the interfacial conditions at the smoothened free surface $y=\myhbar(x,t)$, rather than on the sharp free surface $y=h(x,t)$.  In more detail, in the lubrication limit of the Navier--Stokes equations  we apply the condition of vanishing shear stress and the matching condition for the droplet pressure to equal the capillary pressure at $y=\myhbar(x,t)$.  It can be noted also that  Equations~\eqref{eq:h_reg1} and~\eqref{eq:mob} reduce to the thin-film equation when the convolutions are dropped.

\section{Numerical Simulations}
\label{sec:numerics}

In this section we explore the solutions of Equation~\eqref{eq:h_reg1} using numerical simulations.   For definiteness, we take the filter $K$ to be the inverse of the Helmholtz operator,
\begin{equation}
K*f=(1-\alpha^2\partial_{xx})^{-1}f=\frac{1}{2\alpha}\int_{-\infty}^\infty \mathe^{-|x-y|/\alpha}f(y)\mathd y,
\label{eq:helmholtz}
\end{equation}
for all continuous, integrable functions on the real line. An advantage of working with the Helmholtz kernel is that it confers on $\myhbar(x,t)$ the following property: 

\vspace{0.1in}
\noindent {\textbf{Theorem 3.1:}}  \textit{Under suitable boundary conditions, the integral of the diffuse free-surface height $\myhbar$ is conserved,}
\[
\frac{\mathd}{\mathd t}\int_{-\infty}^\infty \myhbar(x,t)\mathd x=0.
\]

\vspace{0.1in}
\noindent {\textbf{Proof:}}  Starting with Equation~\eqref{eq:h_reg1}, it can be seen that the integral of the bare free-surface height $h(x,t)$ is conserved (subject to appropriate boundary conditions as $|x|\rightarrow \infty$), since the equation for $h_t$ is written in conservative form.  However, we for the Helmholtz kernel, we have $\myhbar=(1-\alpha^2\partial_{xx})^{-1}h$, hence
\begin{equation}
h=\myhbar-\alpha^2\partial_{xx}\myhbar.
\label{eq:hhbar}
\end{equation}
We integrate Equation~\eqref{eq:hhbar} over the whole real line.  Assuming $\partial_x\myhbar\rightarrow 0$ as $|x|\rightarrow \infty$, we obtain 
\[
\int_{-\infty}^\infty h(x,t)\mathd x=\int_{-\infty}^\infty\myhbar(x,t)\mathd x.
\]
Hence, since the integral of $h(x,t)$ is conserved, it follows that the integral of $\myhbar$ is conserved also, and the result follows. \myqed

\vspace{0.1in}
\noindent 
For the purpose of numerical simulations,  we further solve Equation~\eqref{eq:h_reg1} on a truncated domain $x\in (-L,L)$ with periodic boundary conditions; this mimics an infinite domain for sufficiently large $L$.  The meaning of the Helmholtz kernel~\eqref{eq:helmholtz} in the context of periodic boundary conditions is explained below.

\subsection*{Methodology}

Rather than solving Equation~\eqref{eq:h_reg1} directly with the Helmholtz kernel~\eqref{eq:helmholtz}, we solve the evolution for the smoothened free-surface height $\myhbar(x,t)$:
\begin{equation}
\partial_t \myhbar=-K*\partial_x\bigg\{
\left[(1-\alpha^2\partial_{xx})\myhbar\right]\mu(h,\myhbar)\partial_x \big(K*\partial_{xx}\myhbar \big)\bigg\}.
\label{eq:h_reg2}
\end{equation}
Therefore, we view $\myhbar$ as the dynamical variable to be evolved in time.  This is more appropriate than working with $h$ as the dynamical variable, as $\myhbar$ is smoother; hence the numerical method is more stable than would otherwise be the case.  

The numerical method used herein is  a semi-implicit finite-difference scheme, based on the already-validated method developed elsewhere in a different context by~\'O N\'araigh and Thiffeault~\cite{naraigh2010nonlinear}.  We provide a brief description of this method (and accompanying validations) in what follows.  A more detailed development of the numerical methodology (and a comparison with alternative approaches, for instance, the particle method~\cite{carrillo2019blob}) will be the subject of future work.

As such, we discretize $\myhbar(x,t)$ on a uniform grid in space and time, with $i\in \{0,1,\cdots, N\}$ labelling the discrete spatial grid points and $n$ labelling the discrete temporal grid points.  The spatial grid has a grid spacing $\Delta x$, hence, the spatial grid points are located at $x_i=i\Delta x-L$, with $\Delta x= 2L/N$.
Each partial derivative is discretized using centred finite differences.  Hence, $\vech^n=(\myhbar(x_0,t^n),\cdots,\myhbar(x_N,t^n))^T$ is a column vector, and the corresponding centred difference operators (with periodic boundary conditions) are $N\times N$ square matrices, denoted here (in an obvious notation) as $\Diff_1$, and $\Diff_2$.  Hence, the discretized convolution operator $K$ is itself a matrix, $K=(\mathbb{I}_{N\times N}-\alpha^2\Diff_2)^{-1}$.  In this way, we discretize Equation~\eqref{eq:h_reg2} in the temporal domain as follows:
\begin{equation}
\frac{\vech^{n+1}-\vech^n}{\Delta t}=-K\Diff_1
\bigg\{
\left[(1-\alpha^2\Diff_2)\vech^n\right]
\bullet (\vech^n)\bullet (\vech^n)\bullet
(\Diff_1 K \Diff_2 \vech^{n+1})\bigg\},
\label{eq:h_reg_num}
\end{equation}
where the $\bullet$ denotes pointwise multiplication of vectors, and $K\Diff_1$ etc. denote standard matrix products.  Equation~\eqref{eq:h_reg_num} can be re-arranged as
$M \vech^{n+1}=\vech^n$, where $M$ is an $N\times N$ (invertible) square matrix.  Thus, the numerical method is semi-implicit, and $\vech^{n+1}$ is extracted from $\vech^n$ by a matrix inversion at each timestep.  The semi-implicit treatment stabilizes the numerical method and allows for a larger timestep than would otherwise be the case~\cite{naraigh2010nonlinear} (the corresponding explicit method involves a fourth-order diffusion operator, which places severe constraints on the timestep for numerical stability).

\subsection*{Validation}

We use the model initial condition
\begin{equation}
h_0(x)=1+\epsilon\cos(kx),\qquad k=n(2\pi/L),\qquad n\in \{1,2,\cdots\}
\label{eq:linear}
\end{equation}
to validate the numerical method.  Here, $\epsilon\ll 1$ is a small positive parameter.  Physically, this corresponds to a flat interface which is perturbed by a sinusoidal disturbance.  This is realistic in the context of either Equation~\eqref{eq:thif_basic} or~\eqref{eq:h_reg2}, as in such a scenario, the initial condition~\eqref{eq:linear} corresponds to damped capillary waves~\cite{oron1997long}.  As such, we substitute Equation~\eqref{eq:linear} into Equation~\eqref{eq:h_reg2} and expand the solution $\myhbar(x,t)=1+\epsilon\myhbar_1(x,t)+\cdots$ in powers of $\epsilon$, keeping only leading-order terms in $\epsilon$.  The result is
\begin{equation}
\frac{\partial\myhbar_1}{\partial t}=
-K*\partial_x(\partial_x K*\partial_{xx}\myhbar_1)
=
-\,\partial_{xxxx}(K*(K*\myhbar_1)),
\label{eq:linear_pde}
\end{equation}
which is a linear partial differential equation.  We substitute the normal-mode solution $\myhbar_1(x,t)=\mathe^{\imag kx+\sigma t}$ into Equation~\eqref{eq:linear_pde} to produce the dispersion relation
\begin{equation}
\sigma(k)=-\frac{k^4}{(1+\alpha^2 k^2)^2}
\label{eq:disp}
\end{equation}

\begin{figure}
	\centering
		\includegraphics[width=0.6\textwidth]{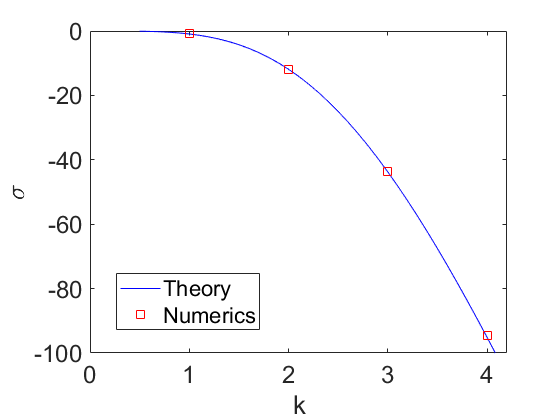}
		\caption{Validation of the numerical method~\eqref{eq:h_reg_num}.  Model parameters: $\alpha=0.2$, $L=2\pi$, $\epsilon=10^{-3}$. Simulation parameters: $\Delta t=10^{-3}$, $N=300$.}
	\label{fig:check_lsa}
\end{figure}

Motivated by the exact solution encoded in Equation~\eqref{eq:disp}, we substitute the initial 
condition~\eqref{eq:linear} into the full nonlinear numerical partial differential equation~\eqref{eq:h_reg_num} and examine the resulting time evolution from the numerical simulation, for a range of values of the wavenumber $k$.   For each considered value of $k$, we monitor the disturbance $\Delta(t)=\|\myhbar(x,t)-1\|_\infty$.  The result of the numerical simulations is fitted to an exponential decay law $\Delta (t)\propto \mathe^{-s_k t}$, where $s_k$ is a fitting parameter, different for each wavenumber $k$.  The values of $s_k$ are tabulated and the results shown in Figure~\ref{fig:check_lsa}.  It can be seen that the dispersion relation thus generated for the numerical results agrees exactly with the analytical dispersion relation in Equation~\eqref{eq:disp}, thereby confirming the correctness of our numerical methods.

\subsection*{Results}

We solve Equation~\eqref{eq:h_reg_num} with the initial condition 
\begin{equation}
\myhbar(x,t=0)=\tfrac{1}{2}h_0\int_{-y_0}^{y_0} \mathe^{-|x-y|/\alpha}(y_0^2-y^2)\mathd y,
\label{eq:ic}
\end{equation}
with $y_0=0.5$ and $h_0=3$.  The effect of different initial conditions has been investigated.  Specifically, we have also looked at Gaussian initial conditions and a piecewise-defined initial condition, with $\myhbar(x,t=0)=(1/2)h_0(y_0^2-x^2)$ inside $|x|<y_0$ and $\myhbar(x,t=0)=0$ outside.  We thereby confirm that the following results are robust with respect to the choice of initial conditions.  We have also carefully tested the  results for numerical convergence: a convergence study is provided in Appendix~\ref{sec:cvg}.

\begin{figure}
	\centering
		\subfigure[]{\includegraphics[width=0.45\textwidth]{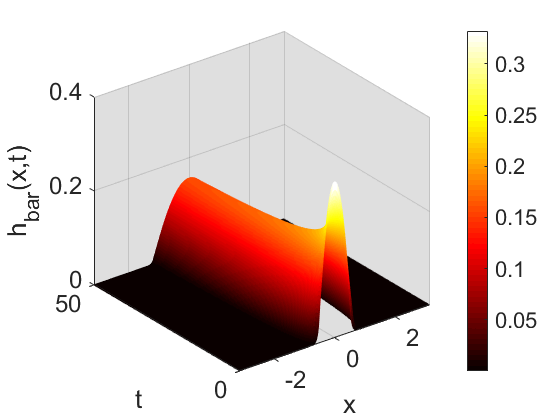}}
		\subfigure[]{\includegraphics[width=0.45\textwidth]{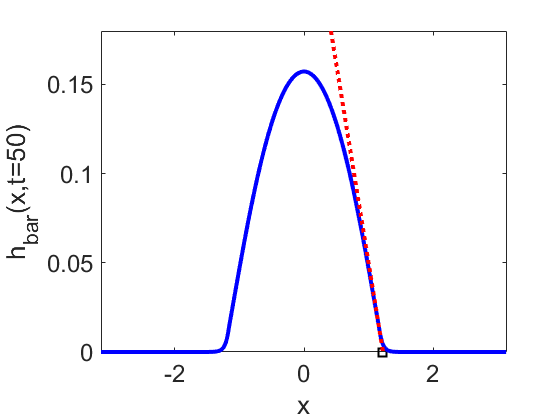}}
		\caption{(a) Spacetime diagram showing the evolution of the diffuse surface height $\myhbar(x,t)$.  (b)  Snapshot of the free-surface height $\myhbar(x,t)$ at $t=50$.  The snapshot also shows the location of the macroscopic contract line $\xmac$.  Model parameter: $\alpha=0.05$.  Numerical parameters: $L=2\pi$, $N=500$ gridpoints, $\Delta t=10^{-2}$.}
	\label{fig:free_surface_evolution}
\end{figure}

\begin{figure}
	\centering
		\includegraphics[width=0.6\textwidth]{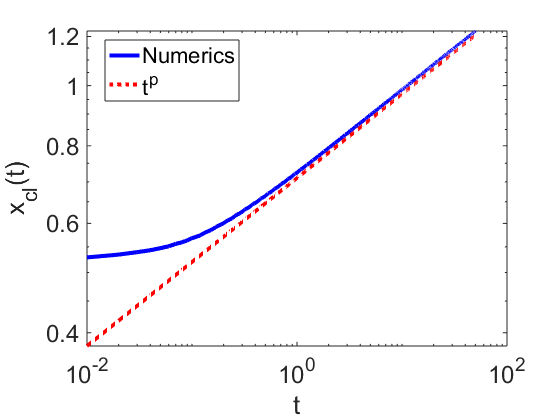}
		\caption{Contact-line evolution based on the numerical simulation, showing a power-law behaviour at late times $\xmac(t)\sim t^p$, witih $p=0.135$.}
	\label{fig:contact_line_evolution}
\end{figure}

Sample numerical results are now shown in Figure~\ref{fig:free_surface_evolution}.  In Figure~\ref{fig:free_surface_evolution}(a) we demonstrate spacetime plot of the diffuse free surface height $\myhbar(x,t)$ as it evolves in space and time.  The lateral extent of the region where $\myhbar$ is significantly different from zero (i.e. the droplet) spreads over time.  Figure~\ref{fig:free_surface_evolution}(b) shows a snapshot of the free-surface profile at $t=50$.  For the purposes of investigation of the numerical results, the macroscopic contact line $\xmac(t)$ is defined operationally.  As such, $\xmac(t)$ is taken to be the realization of the maximum 
\begin{equation}
\max_{x\in[-L,L]} [-\partial_x\myhbar(x,t)].
\label{eq:opdef}
\end{equation}
  The corresponding tangent line is also shown in the figure.
The time evolution of the contact line $\xmac(t)$ is shown in Figure~\ref{fig:contact_line_evolution}.  It can be seen that $\xmac(t)$ behaves as a power law at late times, with $\xmac(t)\sim t^p$ and $p\approx 0.135$, which is obtained by least-squares fitting.   This is very close to the theoretical value $p=1/7$ given by Tanner's Law.

\begin{figure}
	\centering
		\includegraphics[width=0.6\textwidth]{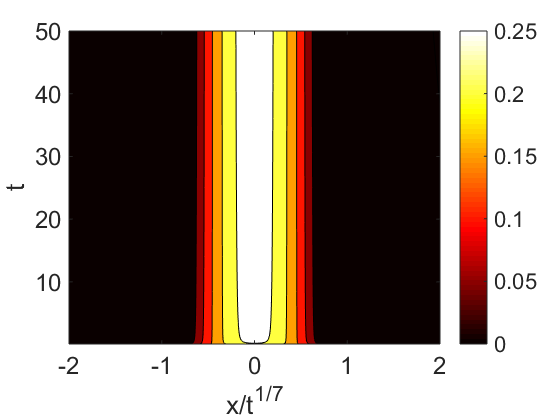}
		\caption{Spacetime diagram in similarity variables showing the evolution of the diffuse surface height 
		$t^{-1/7}\myhbar$}
	\label{fig:self_similar_spacetime}
\end{figure}

We next examine the structure of the numerical solution.  Figure~\ref{fig:self_similar_spacetime} shows a space-time plot of the solution, this time in similarity variables, with $t^{-1/7}\myhbar$ plotted on the $z$-axis, using the scaled spatial variable $\eta=x/t^{1/7}$ (the third dimension along the $z$-axis is shown via a contourplot).  As such, after transient effects have died away, when viewed on the scale of the computational domain, the solution of the regularized model~\eqref{eq:h_reg2} relaxes to a self-similar functional form.

\begin{figure}
	\centering
		\includegraphics[width=0.6\textwidth]{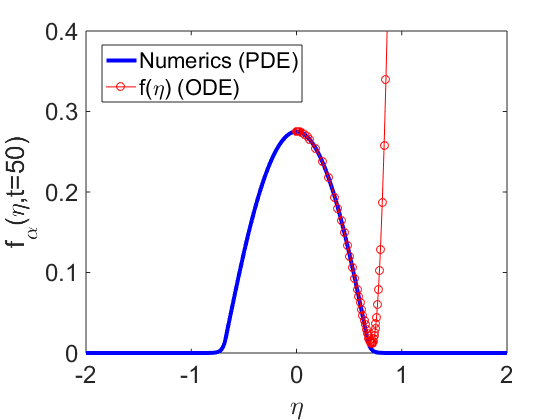}
		\caption{Comparison of the solution of the regularized problem~\eqref{eq:h_reg2} in similarity variables at $t=50$ with the numerical solution of the unregularized problem $f^2f'''=\eta f/7$.  Unadorned solid line: regularized problem.  Line with circles: unregularized problem.}
	\label{fig:pde_ode_compare}
\end{figure}

To understand the results in Figure~\ref{fig:self_similar_spacetime}, a plot of 
$f_{\alpha}(\eta,t):=t^{-1/7}\myhbar(\eta t^{1/7},t)$, with $\eta t^{1/7}=x$ is shown in Figure~\ref{fig:pde_ode_compare}, for $t=50$.  The numerical solution is compared with a solution of the ordinary differential equation $f^2 f'''=\eta f/7$, which is the (singular) similarity equation for the un-regularized dynamics~\eqref{eq:thif_basic}.  The ordinary differential equation is seeded with the initial condition $f'(0)=0$; the additional required initial conditions on $f(0)$ and $f''(0)$ are fed in from the numerical solution of the partial differential equation; specifically, $f(0)=f_\alpha(0,t=50)$, and $f''(0)=f_\alpha(0,t=50)$.  It can be seen that the profiles of $f_\alpha(\eta,t)$ and $f(\eta)$ agree for $|\eta|\ll 1$.  Once the macroscopic contact line at $\eta\approx 1$ is reached, the singular nature of the solution of the un-regularised problem becomes apparent, and $f(\eta)$ diverges. One may take this as an equivalent definition of the mascroscopic contact line, i.e. equivalent to the operational definition~\eqref{eq:opdef}.  In contrast, it is precisely in this region where the smooth nature of the solution of the regularized problem begins to appear, and $f_\alpha(\eta,t)$ tends to zero as $|\eta|\rightarrow \infty$.

\begin{figure}
	\centering
		\includegraphics[width=0.6\textwidth]{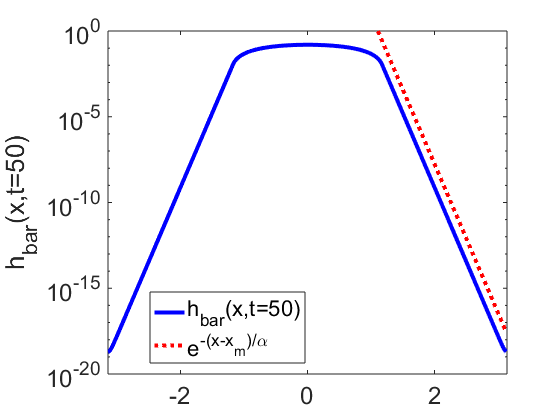}
		\caption{Plot of $\myhbar(x,t=50)$ showing the spatial structure of the solution in the tail, for $|x|\gg \xmac$.}
	\label{fig:hbarstruct}
\end{figure}

To understand the far-field structure of the diffuse free-surface height, we plot in Figure~\ref{fig:hbarstruct} the numerical value of $\myhbar(x,t=50)$, on a semilogarithmic scale.  
The tail of the profile shows a clear exponential decay $\myhbar(x,t)\sim \mathe^{-|x-\xmac|/\alpha}$.  Indeed, it is clear from Figure~\ref{fig:pde_ode_compare} and~\ref{fig:hbarstruct} that the late-time solution of the regularised model~\eqref{eq:h_reg2} with the smoothing kernel~\eqref{eq:helmholtz} is a patchwork of two distinct types:
\begin{equation}
\myhbar(x,t)\sim \begin{cases}
t^{-1/7}f(x/t^{1/7}),\qquad x\ll \xmac(t),\\
A(t)\mathe^{-|x-\xmac(t)|/\alpha},\qquad |x|\gg \xmac(t),
\end{cases}
\end{equation}
with the solution judiciously matching between the two extremes such that $\xmac(t)\sim t^{1/7}$, in agreement with Tanner's Law.

\subsection*{Discussion}

\begin{figure}
	\centering
		\includegraphics[width=0.6\textwidth]{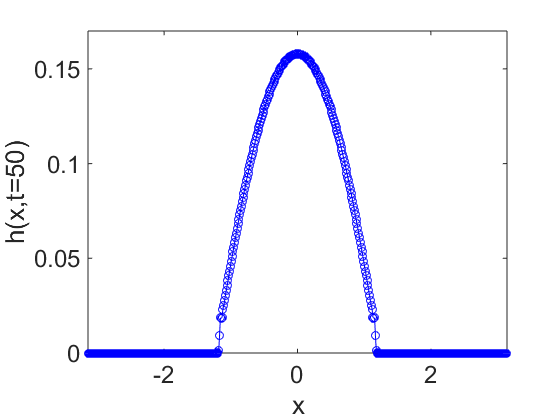}
		\caption{Snapshot of the sharp free-surface height $h(x,t)$ at $t=50$.  Numerical parameters as before.}
	\label{fig:rho_t50}
\end{figure}

The numerical solutions use $\myhbar$ as a dynamical variable, the variable $h=(1-\alpha^2\partial_x^2)\myhbar$ therefore plays a passive role.  A snapshot of $h$ for the above numerical parameters is shown in 
Figure~\ref{fig:rho_t50}.  From the figure, it is apparent that $h=0$ for $|x|\gg \xmac(t)$, this is consistent with $\myhbar(x,t)\sim A(t) \mathe^{-|x-\xmac(t)|/\alpha}$ for $|x|\gg \xmac(t)$, where $A(t)$ is a time-dependent prefactor.  We have deliberately retained the numerical gridpoints in Figure~\ref{fig:rho_t50}: this shows an apparent jump discontinuity in $h(x,t)$ near the position  of the macroscopic contact line.
The presence of a jump discontinuity in $h(x,t)$ is consistent with the moving contact line.    The validity of this claim is demonstrated by the following theorem:

\vspace{0.1in}
\noindent {\textbf{Theorem 3.2:}} 
Consider the solution $h(x,t)$ to Equation~\eqref{eq:h_reg1}.  Regarding the spatial variable $x$, if $h(x,t)$ is in the function class $C^0(-\infty,\infty)$ and piecewise differentiable on $(-\infty,\infty)$, then there is no moving-contact-line solution to Equation~\eqref{eq:h_reg1}.

\vspace{0.1in}
\noindent {\textbf{Proof:}} 
We suppose  that there is a moving contact-line solution $h(x,t)=\phi(x,\xcl) H(x-\xcl)H(x+\xcl)$ to Equation~\eqref{eq:h_reg1}, 
where $\xcl(t)$ is the microscopic contact line, i.e. the minimum positive value of $x$ for which $h(\xcl,t)=0$.  Here also, $H(s)$ denotes the Heaviside function, and $\phi(x,\xcl)$ is a differentiable function on the interval $(-\xcl,\xcl)$, which by the assumption of the moving-contact-line solution satisfies
\[
\Phi(\xcl):=\frac{\partial\phi}{\partial \xcl}\big|_{x=\xcl},\qquad \Phi(\xcl)\neq 0.
\]   
We apply this solution to Equation~\eqref{eq:h_reg1} and integrate from $x=-\infty$ to $x=\infty$.  We obtain
\[
\Phi(\xcl)(\mathd \xcl/\mathd t)=\int_{-\infty}^\infty \frac{\partial}{\partial x}\left[h\mu(h,\myhbar)K*\partial_{xx}\left(K*h\right)\right]\mathd x.
\]
If $h\in C^0(-\infty,\infty)$ in the spatial variable and if $h$ is piecewise differentiable also in the spatial variable, then the integral on the right-hand side can be broken up into different parts and evaluated to give zero, giving $\Phi(\xcl)(\mathd \xcl/\mathd t)=0$, hence $\mathd \xcl/\mathd t=0$, hence no moving contact line. \myqed

\vspace{0.1in}
\noindent Therefore, we conclude that the moving contact line in the numerical simulations corresponds to a sharp free-surface height $h(x,t)$ which is piecewise differentiable, but with jump discontinuities.   As such, the sharp free-surface height $h(x,t)$ satisfies Equation~\eqref{eq:h_reg1} in a weak sense, with $h\in C^{-1}(-\infty,\infty)$ (i.e. $h$ possesses a finite number of jump discontinuities).  Using the convolution~\eqref{eq:helmholtz}, we can conclude that $\myhbar\in C^1(-\infty,\infty)$.  We furthermore look at the convolution of Equation~\eqref{eq:h_reg1} (i.e. Equation~\eqref{eq:h_reg2}) -- this is the equation satisfied by $\myhbar$.  By counting derivatives on both sides, Equation~\eqref{eq:h_reg2} is consistent with $\myhbar\in C^1(-\infty,\infty)$, indicating that $\myhbar$ satisfies Equation~\eqref{eq:h_reg2}  in a strong sense.  We emphasize that although this discussion is consistent with the very precise numerical simulations carried out herein, the rigorous proof that $\myhbar \in C^1(-\infty,\infty)$ has not yet been established. This proof will be the subject of future work.

\section{Discussion and Conclusions}
\label{sec:conc}

In summary, we have introduced a regularsed thin-film equation which describes contact-line motion.  The method does not rely on a slip length or a precursor film.  The method is inspired by the diffuse-interface concept and it involves a smoothened or diffuse free-surface profile $\myhbar(x,t)$.  However, the method still contains a sharp interface, which can be obtained via  deconvolution.  The method reproduces Tanner's law for droplet spreading.  Based on the numerical results and on counting the derivatives in the regularized thin-film equation, the diffuse profile $\myhbar$ gives rise to a strong solution of the thin-film equation.  However, this should be checked rigorously using theoretical methods (e.g. along the lines of Reference~\cite{bernis1990higher}).  The model in its present guise can also be used as a description for spreading over heterogeneious surfaces, by introducing a spatial dependence into the lengthscale $\alpha$, e.g. letting $K*f=\{1-\partial_x(\alpha^2(x)\partial_x ]\}^{-1}f$ in Equation~\eqref{eq:helmholtz}.

The model as formulated currently allows only for indefinite droplet spreading, corresponding to spreading on a hydrophobic surface.  In order to allow for arrested spreading (and hence, a static contact angle), the model will require the introduction of extra physics, for instance, by adding a body-force potential of the Van der Waals type to Equation~\eqref{eq:h_reg1}.  Equally, the model may be extended beyond the limit of lubrication theory, by combining the theoretical arguments in Section~\ref{sec:theory} with the general level-set formulation of two-phase flow.  In this way, it is hoped that the present relatively simple model can serve as a template for geometric diffuse-interface methods for general two-phase flows.

\subsection*{Acknowledgements}

This work has been produced as part of ongoing work within the ThermaSMART network.  The ThermaSMART network has received funding from the European Union's Horizon 2020 research and innovation programme under the Marie Sklodowska--Curie grant agreement No. 778104. DH was partially supported by EPSRC Standard [Grant number EP/N023781/1], entitled, ``Variational principles for stochastic parameterisations in geophysical fluid dynamics". CT acknowledges support from the Alexander von Humboldt Foundation (Humboldt Research Fellowship for Experienced Researchers) as well as from the German Federal Ministry for Education and Research.
LON further acknowledges the UCD Research Sabbatical Leave for Faculty scheme, as well as helpful discussions with Vakhtang Putkaradze.


\appendix

\section{Convergence analysis}
\label{sec:cvg}

\begin{figure}
	\centering
		\subfigure[]{\includegraphics[width=0.45\textwidth]{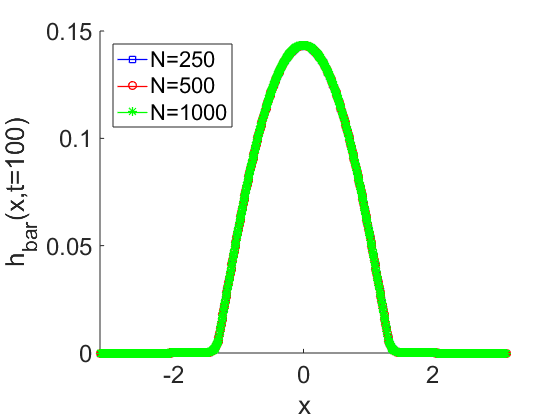}}
		\subfigure[]{\includegraphics[width=0.45\textwidth]{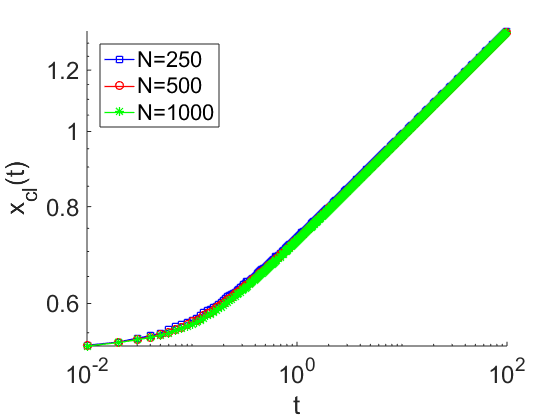}}
		\caption{Convergence study: effect of varying $N$ at fixed $\Delta t=10^{-2}$.  The snapshots in (a) are taken at $t=100$.}
	\label{fig:cvg1}
\end{figure}

\begin{figure}
	\centering
		\includegraphics[width=0.6\textwidth]{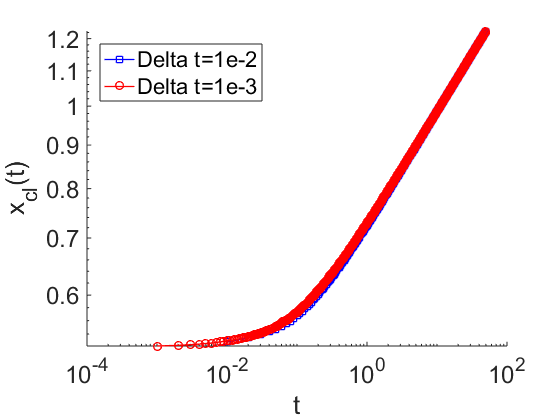}
		\caption{Convergence study: effect of varying $\Delta t$ on the plot of the macroscopic contact line position $\xmac(t)$}
	\label{fig:cvg2}
\end{figure}

In this Appendix, we look at the convergence of the numerical method for the base case considered in Section~\ref{sec:numerics} with $\alpha=0.05$, $L=2\pi$, $N=500$, and $\Delta t=10^{-2}$.  We show the effect of varying the number of grid points $N$ and the timestep $\Delta t$.  In this way, we demonstrate that the numerical results shown in Section~\ref{sec:numerics} are converged.
As such, the structure of $\myhbar(x,t)$ is shown in Figure~\ref{fig:cvg1}(a) for $\Delta t=10^{-2}$, $t=100$, and various values of $N$.  There is no visible change in the structure of $\myhbar(x,t)$ when $N$ is varied between $250$ and $1000$.  Similarly, the position of the macroscopic contact line $\xmac(t)$ is plotted in 
Figure~\ref{fig:cvg1}(b) for the various values of $N$ between $250$ and $1000$.  There is little or no difference between the different plots of $\xmac(t)$ versus $t$ for the various values of $N$.
In Figure~\ref{fig:cvg2} we further show the time evolution of  $\xmac(t)$ for fixed $N=500$ and various values of $\Delta t$.  Again, there is little or no difference between the different plots of $\xmac(t)$ showing that the numerical results presented in the main paper are converged.

Finally, it can be noted that the convergence of the numerical method is rather senstive to the choice of mobility.  For instance, using $\overline{\mu}=h^2$ rather than Equation~\eqref{eq:mob} leads to non-convergent results.  The choice $\overline{\mu}=h^2$ corresponds  to the Navier--Stokes equations in the lubrication limit with a regularized pressure $p=-K*\partial_{xx}(K*h)$ but the application of the no-stress boundary condition $\partial u/\partial y=0$ on $y=h$ rather than on $y=\myhbar$.  The non-convergence of the numerical results in this instance underlines the importance of using the diffuse-interface $\myhbar$ consistently in the formulation of the interfacial stress conditions; it also underlines the importance of choosing a mobility function with a physical rationale.

\end{document}